\documentclass[prd,floatfix,amsmath,nofootinbib,amssymb,floatfix]{revtex4}

\usepackage{amsmath}
\usepackage{amssymb}
\usepackage{graphicx}
\usepackage{float}
\usepackage{bm}
\usepackage{url}

\usepackage[colorlinks,
            citecolor=blue,
            anchorcolor=red,
            menucolor=red,
            linkcolor=red,
            filecolor=red,
            runcolor=red,
            urlcolor=blue,
            frenchlinks=red]{hyperref}

\newcommand{\sumint}{\kern 0.2 em {\textstyle\sum} \kern -1.1 em \int_X}

\begin{document}

\title{The cos$\phi_R$ asymmetry of dihadron production in double longitudinally polarized SIDIS}
\author{Wei Yang}
\affiliation{School of Physics, Southeast University, Nanjing 211189, China}
\author{Hui Li}
\affiliation{School of Physics, Southeast University, Nanjing 211189, China}
\author{Zhun Lu}
\email{zhunlu@seu.edu.cn}
\affiliation{School of Physics, Southeast University, Nanjing 211189, China}

\begin{abstract}
We present a study on the double longitudinal-spin asymmetry of dihadron production in semi-inclusive deep inelastic scattering, in which the total transverse momentum of the final-state hadron pairs is integrated out.
In particular, we investigate the origin of the $\cos\phi_R$ azimuthal asymmetry for which we take into account the coupling of the helicity distribution $g_1$ and the twist-3 dihadron fragmentation function $\widetilde{D}^{\sphericalangle}$. We calculate the $s$-wave and $p$-wave interference term $\widetilde{D}^{\sphericalangle}_{ot}$ in a spectator model. We estimate the $\cos\phi_R$ asymmetry at the kinematics of COMPASS which is collecting data on dihadron production in polarized deep inelastic scattering. The prediction of the same asymmetry at JLab 12GeV and a future EIC are also presented. Our study indicates that measuring the $\cos\phi_R$ asymmetry in SIDIS may be a ideal way to probe the dihadron fragmentation function $\widetilde{D}^{\sphericalangle}$.
\end{abstract}
\maketitle
\section{Introduction}
\label{Sec.introduce}

The spin and azimuthal asymmetries in semi-inclusive deep inelastic scattering (SIDIS) process~\cite{Mulders:1995dh,Kotzinian:1994dv,Boer:1997nt,Goeke:2005hb,Bacchetta:2006tn} have been recognized as useful tools for exploring both the partonic structure of the nucleon and the hadronization mechanism of hadrons, which are among the main tasks in QCD and hadronic physics.
According to factorization~\cite{Ji:2004wu}, the cross-section of SIDIS process can be expressed as the convolution of the parton distribution function, the fragmentation function, and the hard scattering factor.
Distribution and fragmentation functions are important nonperturbative quantities encoding the internal parton and spin structure of nucleon as well as the fragmentation mechanism.

Recently, a lot of attention has also been paid on the higher-twist contributions~\cite{Efremov:2002ut,Gamberg:2003pz,Avakian:2007mv,Mao:2012dk,
Mao:2014aoa,Mao:2014fma,Song:2010pf,Wei:2016far,Yang:2016mxl,Lorce:2016ugb,Bastami:2018xqd} in SIDIS.
Although the rigorous proof on factorization at the twist-3 level in SIDIS has not been achieved~\cite{Bacchetta:2008xw,Gamberg:2006ru}, it is shown~\cite{Bacchetta:2006tn,Bacchetta:2003vn} that those effects are related to the twist-3 distributions and fragmentation functions based on the tree-level calculation: the spin or azimuthal dependent structure function can be expressed as the convolution of the twist-3 distribution/fragmentation functions and the twist-2 fragmenation/distribution functions.
Moreover, there are existing and ongoing experimental measurements on the azimuthal asymmetries of single-hadron and hadron pair (dihadron from a single jet) production at higher-twist level in polarized SIDIS by COMPASS, HERMES, and CLAS at JLab, not only for single hadron production~\cite{Airapetian:2006rx,Airapetian:2005jc,Alekseev:2010dm,Aghasyan:2011ha,Gohn:2014zbz,Airapetian:2019mov}, but also for dihadron production~\cite{S.Sirtl}.
For the later case, the azimuthal asymmetry at twist-2 level has been measured and was applied
to extract~\cite{Bacchetta:2011ip,Bacchetta:2012ty,Radici:2015mwa,Radici:2018iag} transversity from SIDIS data~\cite{Airapetian:2008sk,Adolph:2012nw,Adolph:2014fjw} based on the coupling of transversity and the chiral-odd dihadron fragmentation function $H_1^{\sphericalangle}$~\cite{Collins:1993kq,Jaffe:1997hf,Bacchetta:2002ux,Boglione:1999pz,Boer:1999mm,Bianconi:1999cd,Radici:2001na,Courtoy:2012ry} in the collinear factorization.

In this work, we study the azimuthal asymmetry of dihadron production in double longitudinally polarized SIDIS $l^\rightarrow+p^\rightarrow \rightarrow l+ h_1+h_2+X$ in the case the total transverse momentum of the dihadron is integrated out.
As shown in Ref.~\cite{Bacchetta:2003vn}, in this particular process, there are two twist-3 terms that might give rise to the asymmetry with a $\cos\phi_R$ modulation.
The first one is the coupling of the T-odd twist-3 distribution $e_L(x)$ and the twist-2 dihadron $H_1^{\sphericalangle}$, while the second one is the combination of the helicity distribution $g_1(x)$ and the twist-3 dihadron fragmentation function (DiFF) $\tilde{D}^{\sphericalangle}$ originating from quark-gluon-quark correlation. Here the symbol $\sphericalangle$ denotes that the corresponding DiFF is the interference fragmentation function.
However, if the time reversal invariance is imposed and the gauge link is the only source of a T-odd distribution, $e_L(x)$ should vanish~\cite{Goeke:2005hb} and the $e_L(x) \,H_1^{\sphericalangle}$ term will not contribute to the $\cos\phi_R$ asymmetry.
Based on this consideration, only one term, the $g_1(x)\,\tilde{D}^{\sphericalangle}$ term, should give rise to the asymmetry.
This is different from the single-hadron production in SIDIS in which usually several twist-3 terms contribute to one observable.
Thus, investigating the $\cos\phi_R$ asymmetry in double polarized SIDIS provides an opportunity to study the unknown twist-3 DiFF  $\tilde{D}^{\sphericalangle}$ in a less ambiguous way.

The remained content of the paper is organized as follows. In Sec.~\ref{Sec.formalism}, we will briefly review the theoretical framework of the $\cos\phi_R$ asymmetry of dihadron production in doubly polarized SIDIS. In Sec.~\ref{Sec.model}, we use a spectator model to calculate the twist-3 dihadron fragmentation function $\widetilde{D}^{\sphericalangle}$. As there is no measurement on $\widetilde{D}^{\sphericalangle}$, model calculation is important for accessing information on this unknown DiFF.
In Sec.~\ref{Sec.numerical}, we numerical estimate the DIFF $\widetilde{D}^{\sphericalangle}$ and the $A^{\cos\phi_R}_{LL}$ double longitudinal-spin asymmetry at the kinematics of COMPASS, JLab 12 GeV as well as EIC. Finally, in Sec.~\ref{Sec.conclusion}, we provide conclusion for this paper.

\section{Formalism of $\cos\phi_R$ asymmetry in dihadron production in SIDIS}
\label{Sec.formalism}

We consider the following dihadron SIDIS process,
\begin{align}
l^{\rightarrow}(\ell)+N^{\rightarrow}(P)\longrightarrow l(\ell')+h_1(P_1)+h_2(P_2)+X,
\end{align}
in which a longitudinally polarized lepton collides on the longitudinally polarized target proton $N$ via the exchange of a virtual photon.
Here the arrow $\rightarrow$ denotes the longitudinally polarization of the beam or the proton target. The corresponding 4-momenta are given in parenthesis in the above formula, then the virtual photon has the momentum $q=\ell- \ell'$.
$P$ is the momentum of the target with mass $M$. In this process, the final-state quark with momentum $k = p + q$ then fragments into two final-state hadrons,
$h^+$ and $h^-$, plus unobserved state X. The momenta of the pair are denoted by $P_1$, $P_2$, respectively.
We present the following kinematical variables that are necessary to describe the differential cross section and express the DiFFs,
\begin{align}
\label{eq:invariants}
&x=\frac{k^+}{P^+},\qquad y=\frac{P\cdot q}{P\cdot l},\qquad z_{i}=\frac{P_{i}^-}{k^-},\\
&z=\frac{P_{h}^-}{k^-}=z_1+z_2,\qquad Q^2=-q^2,\qquad s=(P+l)^2, \\
&P_h=P_1+P_2, \qquad R=(P_1-P_2)/2, \qquad M_h=\sqrt{P_h^2} .
\end{align}
Furthermore, the 4-vectors are given in terms of the light-cone coordinates $a^\mu=(a^+, a^-, a_T)$, where the light-cone components are defined as $a^{\pm}=(a^0\pm a^3)/\sqrt{2}$, and $a_T$ is a bidimensional vector of the transverse component.
Therefore, $x$ represents the lightcone momentum fraction of the initial quark, $z_i$ is the lightcone momentum fraction of hadron $h_i$ found in the fragmented quark.
Finally, $M_h$, $P_h$ and $R$ are the invariant mass, the total momentum and the relative momentum of the hadron pair.

\begin{figure}
  \centering
  \includegraphics[width=0.40\columnwidth]{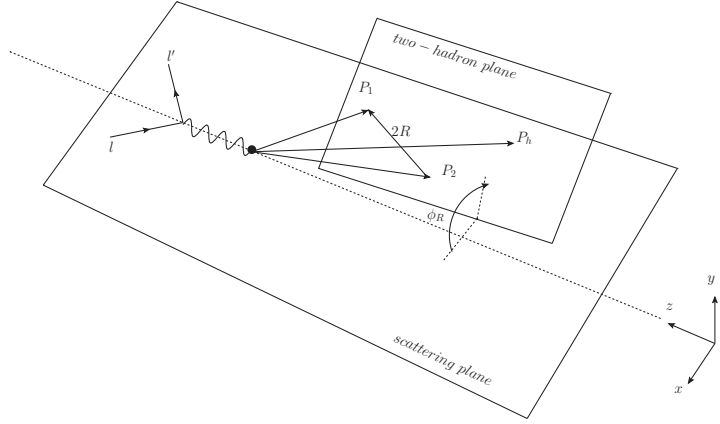}
  \caption{Sketch of the dihadron production in SIDIS process in the lab framel, including the relevant azimuthal angles. The nucleon is assumed to be longitudinally polarized either along or against the direction of the incoming lepton.}
  \label{fig:1}
\end{figure}

We work in the lab frame, in which
the momenta $P_h^{\mu}$, $k^{\mu}$ and $R^{\mu}$ can be decomposed to~\cite{Bacchetta:2006un,Boer:2003cm}
\begin{eqnarray}
P_h^{\mu}&=&\left[P^-_h,\frac{M_h^2}{2P^-_h},\vec{0} \right],\cr
k^{\mu}&=&\left[\frac{P_h^-}{z},\frac{z(k^2+\vec{k}_T^2)}{2P_h^-},\vec{k}_T \right],\cr
R^{\mu}&=&\left[\frac{|\vec{R}|P^-_h}{M_h}\cos\theta,-
\frac{|\vec{R}|M_h}{2P^-_h}\cos\theta,|\vec{R}|\sin\theta\cos\phi_R,|\vec{R}|\sin\theta\sin\phi_R \right].\cr
&=&\left[\frac{|\vec{R}|P^-_h}{M_h}\cos\theta,-\frac{|\vec{R}|M_h}{2P^-_h}\cos\theta,
\vec{R}_T^x,\vec{R}_T^y \right].
\end{eqnarray}
Here, as shown in Fig.~\ref{fig:1}, $\phi_R$ is the angle between the lepton plane and the dihadron plane defined as
\begin{align}
&\cos{\phi_R}= \frac{\hat{q}\times\vec{l}}{|\hat{q}\times\vec{l}|}\cdot\frac{\hat{q}\times\vec{R_T}}
{|\hat{q}\times\vec{R_T}|}, \qquad \sin{\phi_R}= \frac{\vec{l}\times\vec{R_T}\cdot\hat{q}}{|\hat{q}\times\vec{l}||\hat{q}\times\vec{R_T}|}.
\end{align}
where $\hat{q}=\vec{q}/|\vec{q}|$ and $R_T$ is the component of $R$ perpendicular to $P_h$.
The angle $\theta$ is polar angle between the direction of $P_1$ in the center of mass frame of the hadron pair and the direction of $P_h$ in the lab frame.

There are several useful expressions for the scalar products as follows
\begin{align}
&P_h\cdot R= 0,\\
&P_h\cdot k= \frac{M_h^2}{2z}+z\frac{k^2+|\vec{k}_T|^2}{2},\\
&R\cdot k= \left(\frac{M_h}{2z}-z\frac{k^2+|\vec{k}_T|^2}{2M_h}\right)|\vec{R}|\cos\theta-\vec{k}_T\cdot\vec{R}_T .
\end{align}
In particular, there is a relation between the $\vec{R}$ and $M_h$.
\begin{align}
|\vec{R}|=\sqrt{\frac{M^2_h}{4}-m_h^2}.
\end{align}

We will focus on the case the lepton beam and proton target are both longitudinally polarized as well as the case the total transverse momentum are integrated out.
For convenience, in the following formula we introduce the subscript $U$, $L$ for the cross section $\sigma^{}_{XY}$ to indicate the unpolarized or longitudinally polarized states.
Besides, the polarization with respect to the beam and the target are indicated by the first label $X$ and the second label $Y$ of $\sigma^{}_{XY}$, respectively.
We can then summarize the unpolarized and double polarized cross sections as follows Ref.~\cite{Bacchetta:2003vn}:
\begin{align}
\frac{d^6\! \sigma^{}_{UU}}{d\cos \theta\;dM_h^2\;d\phi_R\;dz\;dx\;dy} &=  \frac{\alpha^2}{ Q^2 y}\,\left(1-y+\frac{y^2}{2}\right)   \sum_a e_a^2 f_1^a(x)\, D_1^a\bigl(z, M_{h}^2, \cos \theta\bigr) , \label{eq:crossUU} \\
\frac{d^6\! \sigma^{}_{LL}}{d\cos \theta\;dM_h^2\;d\phi_R\;dz\;dx\;dy} &=  \frac{\alpha^2}{Q^2 y}\, S_L \sum_a e_a^2 \,y(\frac{y}{2}-1) g^a_1(x)D^a_1(z,M_h^2,\cos\theta) \label{eq:crossLL} \nonumber\\
&+\frac{\alpha^2}{Q^2 y}\, S_L \sum_a e_a^2 \,2y\sqrt{2-y}{M\over Q}\frac{|\bm R| }{M_h} \, \cos\phi_{R}\nonumber\\
 &  \times
   \left[xe^a_L(x)H_{1}^{\sphericalangle,a}\bigl(z,M_h^2,\cos\theta\bigr)-\frac{M_h}{Mz} g^a_1(x)\widetilde{D}^{\sphericalangle,a}\bigl(z,M_h^2,\cos\theta\bigr)\right].
\end{align}
In Eq.~(\ref{eq:crossUU}), $f_1^a(x)$ and $D_1^a\bigl(z, M_{h}^2, \cos \theta\bigr)$ denote the unpolarized PDF and unpolarized DiFF for flavor $a$.
The first line in Eq.(~\ref{eq:crossLL}) represents the leading twist contribution, while the second line denotes the twist-3 terms which contributes to the $\cos\phi_R$ asymmetry.
In details, there are two individual contributions that might give rise to the asymmetry.
The first one is $e_L(x)H_{1}^{\sphericalangle}$, in which $e_L(x) = \int d^2 k_T e_L(x,k_T^2)$ is a T-odd twist-3 distribution, and $H_{1}^{\sphericalangle,a}$ is the twist-2 DiFF.
The second one is $g_1(x)\widetilde{D}^{\sphericalangle}$, where $g_1(x)$ is the twist-2 helicity distribution and $\widetilde{D}^{\sphericalangle}$ is the twist-3 DiFF, with the symbol $\sphericalangle$ denoting the interference nature of the DiFF.
However, as shown in Ref.~\cite{Goeke:2005hb}, if $e_L(x,k_T^2)$ only receives contribution from the gauge-link, time-reversal invariance of QCD implies the constraint $\int d^2 k_T e_L(x,k_T^2)=0$, therefore, the contribution $e_L(x)H_{1}^{\sphericalangle}$ vanishes in the collinear limit.
Based on this observation, in this work we only need to consider the $g_1(x)\widetilde{D}^{\sphericalangle}$ term.

The partial-wave analysis of the DiFFs $D_1$ and $\widetilde{D}^{\sphericalangle}$ up to $p$-wave level yields~\cite{Bacchetta:2003vn}:
\begin{align}
&D^a_{1}(z,\cos \theta,M_h^2)=D^a_{1,oo}(z,M_h^2)+D^a_{1,ol}(z,M_h^2)\cos \theta+D^a_{1,ll}(z,M_h^2)(3\cos^2\theta-1),\\
&H_{1}^{\sphericalangle a}(z,\cos \theta, M_h^2)=H_{1,ot}^{\sphericalangle a}(z,M_h^2)+H_{1,lt}^{\sphericalangle a}(z,M_h^2)\cos \theta.\\
&\widetilde{D}^{\sphericalangle}(z,\cos \theta, M_h^2)=\widetilde{D}^{\sphericalangle}_{ot}(z,M_h^2)+\widetilde{D}^{\sphericalangle}_{lt}(z,M_h^2)\cos \theta.
\end{align}
Here, $D^a_{1,oo}(z,M_h^2)$ comes from the pure $s$-wave and $p$-wave contributions, $D^a_{1,ol}$ and $\widetilde{D}^{\sphericalangle}_{ot}(z,M_h^2)$ arise from the interference between a pair in $s$-wave and a pair in $p$-wave.

Following the similar arguments in Ref.~\cite{Bacchetta:2006un}, in this paper we will not consider the $\cos \theta$-dependent terms in the expansion of DiFFs.
When integrating out the angular $\theta$ in the interval $[0, \pi]$ which is our case, the $\cos\theta$-dependent terms should vanish.
Therefore, we focus on the functions $D^a_{1,oo}$ and  $\widetilde{D}^{\sphericalangle}_{ot}$.
In this scenario, the $\cos\phi_R$ asymmetry of dihadron production contribution to the double longitudinally polarized can be expressed as
\begin{align}
A^{\cos\phi_R}_{LL}(x,z,M_h^2) = -\frac{\sum_{a}e^2_a\frac{|\vec{R}|}{Q}
\left[\frac{1}{z} g_1(x)\widetilde{D}^{\sphericalangle}_{ot}(z,M_h^2)\right]}{\sum_{a}e^2_a f^a_1(x) D^a_{1,oo}(z,M_h^2)}.
\label{eq:AcosphiR}
\end{align}
Following the convention used by COMPASS in Ref.~\cite{S.Sirtl},  the depolarization factors are not included in the numerator and the denominator.

\section{Calculation of the DiFF $\widetilde{D}^{\sphericalangle}_{ot}$ in the spectator model}
\label{Sec.model}

The twist-3 DiFF $\widetilde{D}^{\sphericalangle}$ originates from the quark-gluon-quark correlation~\cite{Bacchetta:2003vn},
\begin{eqnarray}
\label{eq:deltaA}
\widetilde{\Delta}_A^{\alpha}(k,P_h,R)&=&\frac{1}{2z}\sum_X\int\frac{d\xi^+d^2\xi_T}{(2\pi)^3}e^{ik\cdot\xi}\langle0|\int_{\pm\infty^+}^{\xi^+}d\eta^+
{\cal{U}}^{\xi_T}_{(\infty^+,\xi^+)}\cr
&&\times g F_\bot^{-\alpha}{\cal{U}}^{\xi_T}_{(\eta^+,\xi^+)}\psi(\xi)|P_h,R;X\rangle\langle P_h,R;X|\bar{\psi}(0){\cal{U}}^{0_T}_{(0^+,\infty^+)}{\cal{U}}^{\infty^+}_{(0_T,\xi_T)}|0\rangle\mid_{\eta^+=\xi^+=0,\eta_T=\xi_T}.
\end{eqnarray}
Here $F_\bot^{-\alpha}$ is the field strength tensor of the gluon. Introducing the covariant derivative $iD^\mu(\xi) = i\partial_\mu + gA_\mu(\xi)$, we can recover also the relation
\begin{eqnarray}
\widetilde{\Delta}_A^{\alpha}(k,P_h,R)&=&{\Delta}_D^{\alpha}(k,P_h,R)-{\Delta}_{\partial}^{\alpha}(k,P_h,R)\,,
\end{eqnarray}
Where
\begin{eqnarray}
{\Delta}_D^{\alpha}(k,P_h,R)&=&z^2\sum_X\int\frac{d\xi^+}{2\pi}e^{ik\cdot\xi}\langle0|{\cal{U}}^{+}_{[0,\xi]}\psi(\xi)iD^\alpha(\xi)
|P_h,R;X\rangle\langle P_h,R;X|\bar{\psi}(0)|0\rangle|_{\xi^-=\xi_T=0}\,,\cr
{\Delta}_{\partial}^{\alpha}(k,P_h,R)&=&z^2 k_T^\alpha \sum_X\int\frac{d\xi^+}{2\pi}e^{ik\cdot\xi}\langle0|{\cal{U}}^{+}_{[0,\xi]}\psi(\xi)
|P_h,R;X\rangle\langle P_h,R;X|\bar{\psi}(0)|0\rangle|_{\xi^-=\xi_T=0}\,.
\end{eqnarray}

After integrating out ${\vec{k}_T}$, we get
\begin{align}
\widetilde{\Delta}_A^{\alpha}(z,\cos \theta,M_h^2,\phi_R)=\frac{z^2|\vec{R}|}{8M_h}\int d^2\vec{k}_T \widetilde{\Delta}_A^{\alpha}(k,P_h,R).
\end{align}
Then, the DiFF $\widetilde{D}^{\sphericalangle}$ can be obtained by the trace
\begin{align}
\frac{R_{T}^{\alpha}}{z}\widetilde{D}^{\sphericalangle}(z,\cos \theta,M_h^2)=4\pi\textrm{Tr}[\widetilde{\Delta}_A^{\alpha}(z,\cos \theta,M_h^2,\phi_R)\gamma^{-}].
\end{align}

Following the approach developed in Ref.~\cite{Bacchetta:2006un}, we will work in the framework of a spectator model for the fragmentation process $q\rightarrow\pi^+\pi^-X$.
Here, the sum over all possible intermediate states $X$ is replaced by an effective on-shell state-the spectator, whose quantum numbers are the same as the initial quark and whose mass is one of the parameters of the model.
The twist-2 DiFFs $D_{1,oo}$ and ${H}^{\sphericalangle}_{1,ot}$ have been studied in Ref.~\cite{Bacchetta:2006un} using the spectator model.
The model was also extended to calculate the twist-3 DiFF $\widetilde{G}^{\sphericalangle}$ in Ref.~\cite{Yang:2019aan}.
In the following, the calculation of the unknown DiFF $\widetilde{D}^{\sphericalangle}_{ot}$ in the same model will be described in details.

\begin{figure}
  \centering
  \includegraphics[width=0.40\columnwidth]{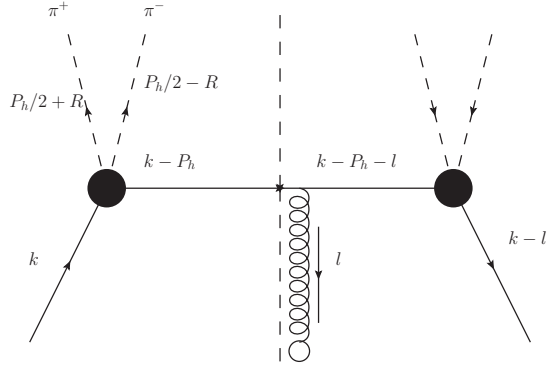}
  \caption{Diagrammatic representation of the correlation function $\widetilde{\Delta}_A^{\alpha}$ in the spectator model.}
  \label{fig:2}
\end{figure}

The corresponding diagram to calculate the quark-gluon-quark correlation for DiFF is shown in Fig.~\ref{fig:2}, in which the left hand side corresponds to the quark-hadron vertex $\langle P_{h},R;X|\bar{\psi}(0)|0\rangle $, and the right hand side corresponds to the vertex  $\langle0|igF_{\perp}^{-\alpha}(\eta^{+})\psi(\xi^{+})|P_{h},R;X\rangle$ which contains gluon rescattering.

As given in details in Ref.~\cite{Bacchetta:2006un}, there are a few prominent channels contributing to the $q\rightarrow\pi^+\pi^-X$ process:

1. $q\rightarrow\pi^+\pi^-X_1$;

2. $q\rightarrow\rho X_2\rightarrow\pi^+\pi^-X_2$;

3. $q\rightarrow\omega X_3\rightarrow\pi^+\pi^-X_3$;

4. $q\rightarrow\omega X'_4\rightarrow\pi^+\pi^-X_4$ with $X_4=\pi^0X'_4$;

5. $q\rightarrow\eta X'_5\rightarrow\pi^+\pi^-X_5$ with $X_5=\chi X'_5$;

6. $q\rightarrow K^0 X_6\rightarrow\pi^+\pi^-X_6$.

In the first process, the quark fragments into an ``incoherent" $\pi^+\pi^-$ pair that are called as ``background"; while in the other five processes the $\pi^+ pi^-$ pair are produced through the decays of the intermediate resonances $\rho$, $\omega$, $\eta$ and $K^0$, responsible for the peaks at $M_h\sim$ 770MeV, 782MeV, 500MeV, and 498MeV, respectively.

As stated in Ref.~\cite{Bacchetta:2006un}, different channels could produce spectators with different masses. For simplicity, here we consider just a single spectator for all channels. The spectator mass is denoted by $M_s$. The choice of using the same spectator for all channels implies in particular that the fragmentation amplitudes of all channels can interfere with each other maximally. In reality, only a fraction of the total events ends up in the same spectator and can thus produce interference effects.

Using the above setup, we can write down the quark-gluon-quark correlator for dihardon fragmentation produced by $s$-wave and $p$-wave
\begin{eqnarray}
\label{Delta1}
&&\widetilde{\Delta}_A^{\alpha}(k,P_h,R)=i\frac{C_F\alpha_s}{2(2\pi)^2(1-z)P_h^-}\frac{1}{k^2-m^2}\int \frac{d^4l}{(2\pi)^4}(l^-g_T^{\alpha\mu}-l^\alpha_T g^{-\mu})\cr
&&\frac{(k\!\!\!/ -l\!\!\!/ +m)(F^{s\star} e^{-\frac{k^2}{\Lambda_s^2}}+F^{p\star} e^{-\frac{k^2}{\Lambda_p^2}} R\!\!\!/)(k\!\!\!/ -P\!\!\!/_h-l\!\!\!/+m_s)\gamma_{\mu}(k\!\!\!/ -P\!\!\!/_h+m_s)(F^{s} e^{-\frac{k^2}{\Lambda_s^2}}+F^{p} e^{-\frac{k^2}{\Lambda_p^2}} R\!\!\!/)(k\!\!\!/+m)}{(-l^-\pm i\epsilon)((k-l)^2-m^2-i\epsilon)((k-P_h-l)^2-m_s^2-i\epsilon)(l^2-i\epsilon)}.
\end{eqnarray}
Here, $m$ and $m_s$ are the mass of the quark and the spectator, respectively, and
$\Lambda_{s}$ and $\Lambda_{p}$ are the cutoffs for the quark momentum with the form~\cite{Bacchetta:2006un}
\begin{align}
\Lambda_{s,p} = \alpha_{s,p} z^{\beta_{s,p}} (1-z)^{\gamma_{s,p}},
\end{align}
The exponential form factors are introduced to suppress the contributions from higher quark virtuality~\cite{Gamberg:2003eg}.
The open circle in Fig.~\ref{fig:2} denote The factor $(l^-g_T^{\alpha \mu} -l_T^\alpha g^{-\mu})$ coming form the Feynman rule for the gluon field strength tensor in the operator definition of the correlator~\ref{eq:deltaA}.
In addition, $F^s$ and $F^p$ are the quark-dihadron-spectator vertices associated with the $s$-wave and $p$-wave contributions, particular, the $p$-wave vertices contain the real part and image part, which are both contribution to the twist-3 fragmentation function $\widetilde{D}^{\sphericalangle}$, respectively, and in Ref.~\cite{Bacchetta:2006un} they are parameterized as:
\begin{align}
F^s& = f_s \nonumber\,,\\
F^p& = f_\rho \frac{(M_h^2-M_\rho^2)-i\Gamma_\rho M_\rho}{(M_h^2-M_\rho^2)+\Gamma_\rho^2 M_\rho^2}+f_\omega \frac{(M_h^2-M_\omega^2)-i\Gamma_\omega M_\omega}{(M_h^2-M_\rho^2)+\Gamma_\omega^2 M_\omega^2}\nonumber\\
&-if'_\omega \frac{\sqrt{\hat{\lambda}(M_\omega^2,M_h^2,m_\pi^2)}\Theta(M_\omega-m_\pi-M_h)}
{4\pi\Gamma_\omega^2[4M_\omega^2m_\pi^2+\hat{\lambda}(M_\omega^2,M_h^2,m_\pi^2)]^{\frac{1}{4}}}\,.
\end{align}
Here, $\hat{\lambda}(M_\omega^2,M_h^2,m_\pi^2)=(M_\omega^2-(M_h+m_\pi)^2)(M_\omega^2-(M_h-m_\pi)^2)$ and $\Theta$ denotes the unit step function.
The first two terms of $F^p$ can be identified with the contributions of the $\rho$ and the $\omega$ resonances decaying into two pions. The masses and widths of the two resonances are taken from the PDG~\cite{Eidelman}: $M_\rho = 0.776 \mathrm{GeV}$, $\Gamma_\rho = 0.150 \mathrm{GeV}$, $M_\omega = 0.783 \mathrm{GeV}$, $\Gamma_\omega = 0.008 \mathrm{GeV}$.

To identify the contributions from the correlator to $\widetilde{D}^{\sphericalangle}_{ot}$, we expand (\ref{Delta1}) in the following way,
\begin{eqnarray}
\label{eq:Delta}
&&\widetilde{\Delta}_A^{\alpha}(z,\cos \theta,M_h^2,\phi_R)=i\frac{C_F\alpha_s z^2|\vec{R}|}{16(2\pi)^5(1-z)M_h P_h^-}\int d|\vec{k}_T|^2\int d^4l\frac{l^-g_T^{\alpha\mu}-l^\alpha_T g^{-\mu}}{k^2-m^2}\cr
&&\bigg{[}+|F^{s}| e^{-\frac{2k^2}{\Lambda_s^2}}\frac{(k\!\!\!/ -l\!\!\!/ +m)(k\!\!\!/ -P\!\!\!/_h-l\!\!\!/+m_s)\gamma_{\mu}(k\!\!\!/ -P\!\!\!/_h+m_s)(k\!\!\!/+m)}{(-l^-\pm i\epsilon)((k-l)^2-m^2-i\epsilon)((k-P_h-l)^2-m_s^2-i\epsilon)(l^2-i\epsilon)}\cr
&&+|F^{p}| e^{-\frac{2k^2}{\Lambda_p^2}}\frac{(k\!\!\!/ -l\!\!\!/ +m) R\!\!\!/(k\!\!\!/ -P\!\!\!/_h-l\!\!\!/+m_s)\gamma_{\mu}(k\!\!\!/ -P\!\!\!/_h+m_s)R\!\!\!/(k\!\!\!/+m)}{(-l^-\pm i\epsilon)((k-l)^2-m^2-i\epsilon)((k-P_h-l)^2-m_s^2-i\epsilon)(l^2-i\epsilon)}\cr
&&+(F^{s\star}F^{p}) e^{-\frac{2k^2}{\Lambda_{sp}^2}}\frac{(k\!\!\!/ -l\!\!\!/ +m)(k\!\!\!/ -P\!\!\!/_h-l\!\!\!/+m_s)\gamma_{\mu}(k\!\!\!/ -P\!\!\!/_h+m_s) R\!\!\!/(k\!\!\!/+m)}{(-l^-\pm i\epsilon)((k-l)^2-m^2-i\epsilon)((k-P_h-l)^2-m_s^2-i\epsilon)(l^2-i\epsilon)}\cr
&&+(F^{s}F^{p\star}) e^{-\frac{2k^2}{\Lambda_{sp}^2}}\frac{(k\!\!\!/ -l\!\!\!/ +m)R\!\!\!/(k\!\!\!/ -P\!\!\!/_h-l\!\!\!/+m_s)\gamma_{\mu}(k\!\!\!/ -P\!\!\!/_h+m_s) (k\!\!\!/+m)}{(-l^-\pm i\epsilon)((k-l)^2-m^2-i\epsilon)((k-P_h-l)^2-m_s^2-i\epsilon)(l^2-i\epsilon)}\bigg{]}\,.
\end{eqnarray}
where $2/\Lambda^2_{sp}=1/\Lambda^2_{s}+1/\Lambda^2_{p}$.

In the spectator model, the factor $k^2$ in Eq.~(\ref{eq:Delta}), compatible with the on-shell condition of the spectator, is given by
\begin{eqnarray}
k^2&=&\frac{z}{1-z}|\vec{k}_T|^2+\frac{M_s^2}{1-z}+\frac{M_h^2}{z}\,.
\end{eqnarray}

In Eq.~(\ref{eq:Delta}), the first term corresponds to the pure $s$-wave contribution, the second term corresponds to the pure $p$-wave contribution; whereas the third and fourth term correspond to the interference of $s$ and $p$-wave contributions.
According to the partial-wave analysis for $\widetilde{D}^{\sphericalangle}$, one can find that only the third and fourth terms contribute to $\widetilde{D}^{\sphericalangle}_{ot}$.
Integrating over the internal momentum $l$ yields the final expression for $\widetilde{D}^{\sphericalangle}_{ot}(z,M_h^2)$:
\begin{align}
\widetilde{D}^{\sphericalangle}_{ot}(z,M_h^2)& =-\frac{\alpha_s C_F z^2 |\vec{R}|}{4 (2\pi)^4 (1-z)M_h}\int d|\vec{k}_T|^2 e^{-\frac{2 k^2}{\Lambda_{sp}^2}}\frac{1}{k^2-m^2}
\bigg{\{}\textrm{Re}(F^{s*}F^p) C\nonumber\\
&+\textrm{Im}(F^{s*}F^p) m_s \big{[}(k^2-m^2)(A+zB)-2(A k^2 + B P_h\cdot k)\big{]} \bigg{\}}\,, \label{eq:dtilde}
\end{align}
Here, the coefficient $C$ corresponds to real part of the integration over $l$ and has the expression
\begin{align}
&C=m\int_0^1 dx \int_0^{1-x} dy \frac{[x+(1-z)y-2](k^2-2k\cdot P_h-m_s^2+m_h^2)+(x+2y)k\cdot P_h-(x+y)k^2-ym_h^2}{x(1-x)k^2+2k\cdot(k-P_h)xy+m^2 x+m_s^2y+y(y-1)(k-P_h)^2}\,,\label{eq:cterm}
\end{align}
Note that it is proportional to the fragmenting quark mass $m$.
The factor ${A}$ and ${B}$ appear in Eq.~(\ref{eq:dtilde}) have the forms
\begin{align}
A&={I_{1}\over \lambda(m_h,m_s)} \left(2k^2 \left(k^2 - m_s^2 - m_h^2\right) {I_{2}\over \pi}+\left(k^2+m_h^2 - m_s^2\right)\right), \\
B&=-{2k^2 \over \lambda(m_h,m_s) } I_{1}\left (1+{k^2+m_s^2-m_h^2 \over \pi} I_{2}\right)\,.
\end{align}
And the functions $I_{i}$ have the forms~\cite{Amrath:2005gv}
\begin{align}
I_{1} &=\int d^4l \delta(l^2) \delta((k-l)^2-m^2) ={\pi\over 2k^2}\left(k^2-m^2\right)\,, \\
I_{2} &= \int d^4l { \delta(l^2) \delta((k-l)^2-m^2)\over (k-P_h-l)^2-m_s^2}
={\pi\over 2\sqrt{\lambda(m_h,m_s)} }  \ln\left(1-{2\sqrt{ \lambda(m_h,m_s)}\over k^2-m_h^2+m_s^2 + \sqrt{ \lambda(m_h,m_s)}}\right)\,.
\end{align}
with $\lambda(m_h,m_s)=(k^2-(m_h+m_s)^2)(k^2-(m_h-m_s)^2)$.

\section{Numerical Estimate for $\widetilde{D}^{\sphericalangle}_{ot}$ and the double spin asymmetry}
\label{Sec.numerical}

In order to obtain the numerical results of the fragmentation function $\widetilde{D}^{\sphericalangle}_{ot}(z,M_h^2)$, we need to know the values for the model parameters $m_s$, $\alpha_{s,p}$, $\beta_{s,p}$ and $\gamma_{s,p}$. For these we adopt them from Ref.~\cite{Bacchetta:2006un}:
\begin{align}
\alpha_s& = 2.60\pm0.05 \mathrm{GeV}^2\,,\qquad  \beta_s = -0.751\pm0.008\,,\qquad \gamma_s = -0.193\pm0.004 \nonumber\,,\\
\alpha_p& = 7.07\pm0.11 \mathrm{GeV}^2\,,\qquad  \beta_p = -0.038\pm0.003\,,\qquad \gamma_p = -0.085\pm0.004 \nonumber\,,\\
f_s& = 1197\pm2 \mathrm{GeV}^{-1}\,,\qquad  f_\rho = 93.5\pm1.6\,,\qquad f_\omega = 0.63\pm0.03 \nonumber\,,\\
f'_\omega& = 75.2\pm1.2 \,,\qquad  M_s = 2.97\pm 0.04M_h\,.
\end{align}

For the quark mass $m$, we will adopt two different choices for comparison.
The first one is $m=0$ GeV following the adoption in Ref.~\cite{Bacchetta:2006un}. In this choice the $C$ term (Eq.~(\ref{eq:cterm})) vanishes since it is proportional to $m$. Therefore, in this particular case only the term containing $\textrm{IM}(F^{s*}F^p)$ in Eq.~(\ref{eq:dtilde}) contributes to $\widetilde{D}^{\sphericalangle}_{ot}$ numerically.
In the second choice we adopt $m=0.3$ GeV, which is consistent with the value for
the quark massl chosen in Refs.~\cite{Bacchetta:2008af} and \cite{Bacchetta:2007wc}.

\begin{figure}
  \centering
  \includegraphics[width=0.49\columnwidth]{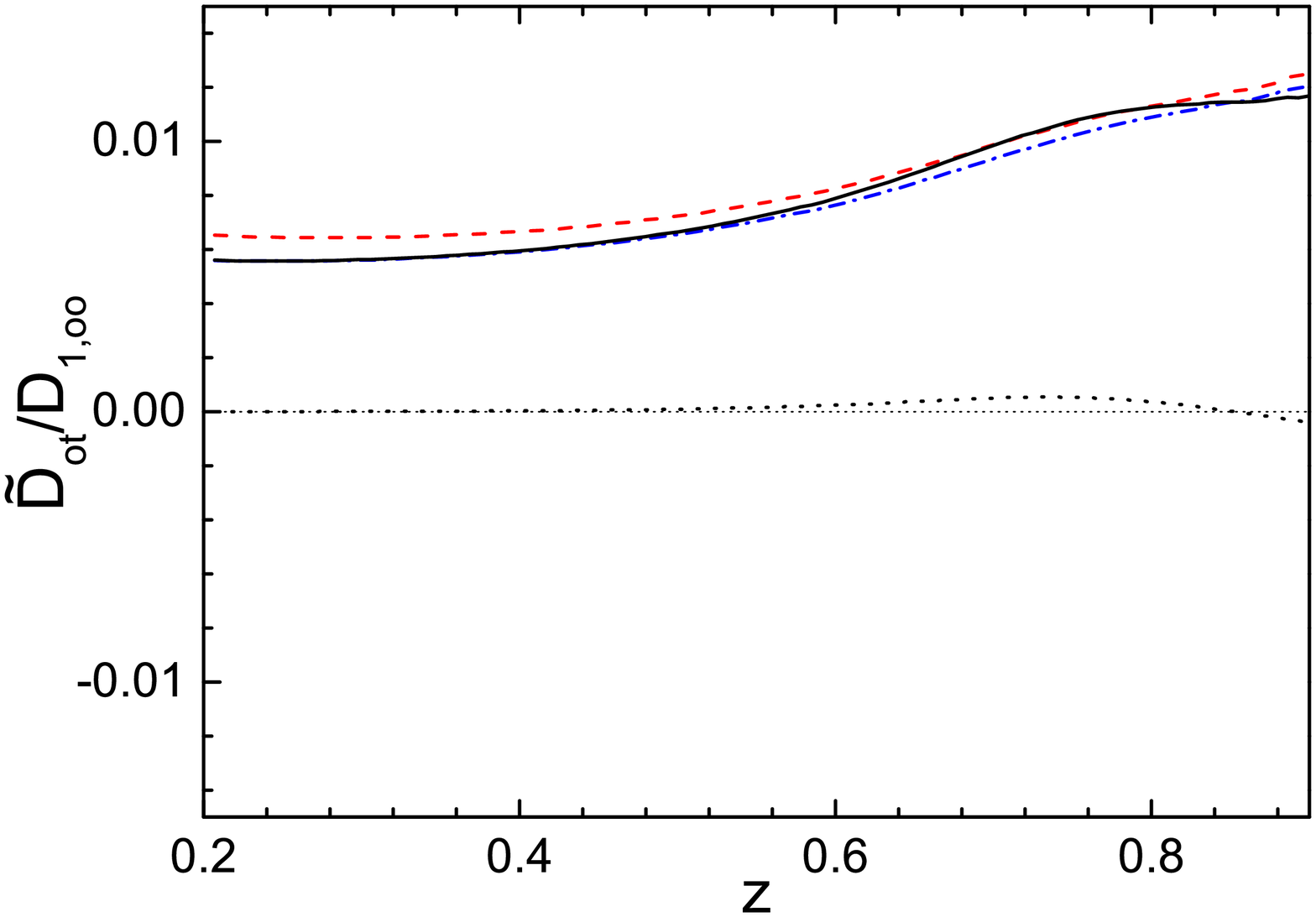}
  \includegraphics[width=0.49\columnwidth]{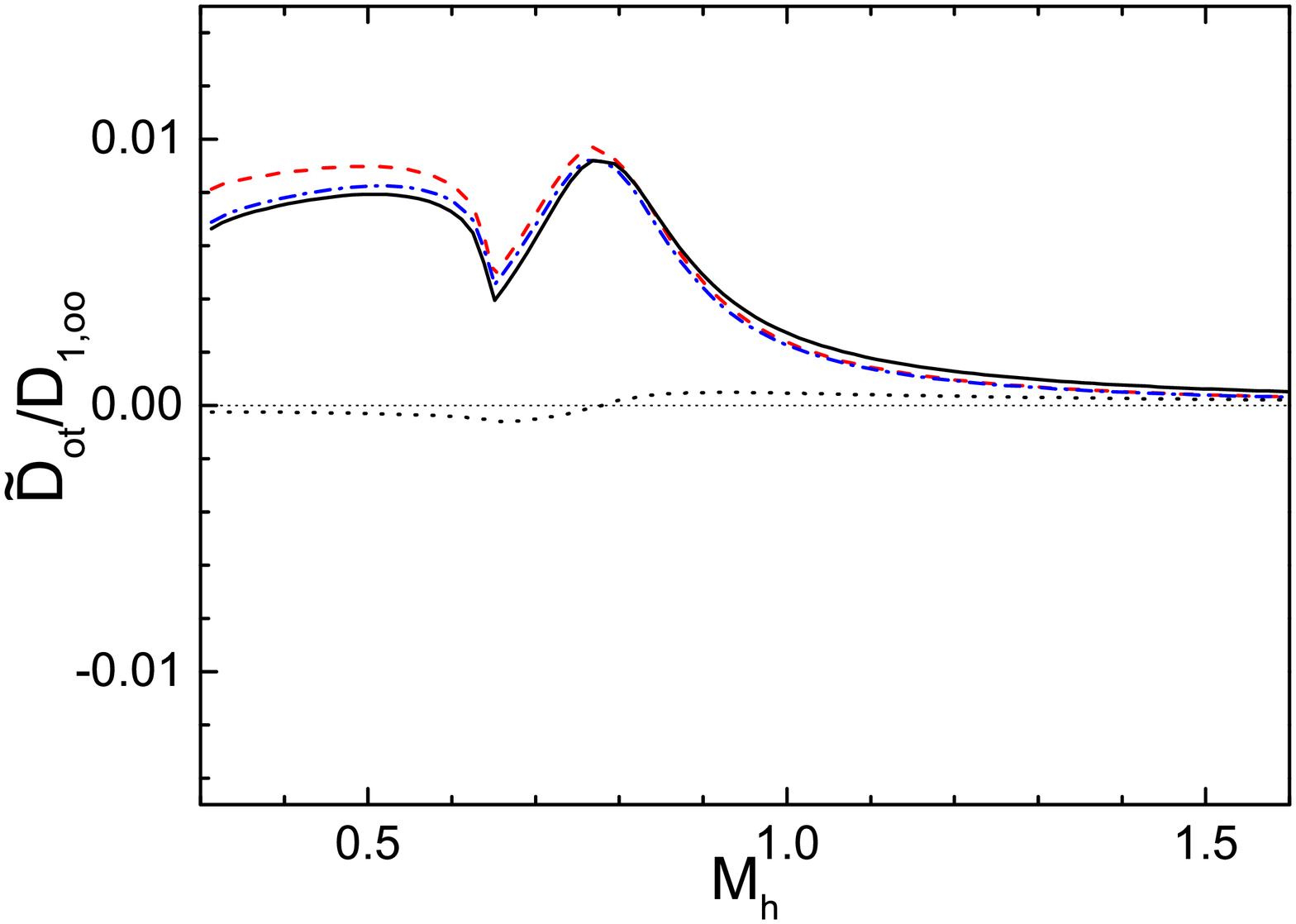}
  \caption{The twist-3 D0iFF $\widetilde{D}^{\sphericalangle}_{ot}$ as the functions of $z$ (left panel) and $M_h$ (right panel) in the spectator model, normalized by the unplarized DiFF $D_{1,oo}$.
The dashed line corresponds to the result for $m=0$ GeV.
The dashed-dotted line corresponds to the result from the Im$(F^{s*} F^p)$ term for $m=0.3$ GeV, while the dotted line corresponds to the result from the $C$ term in Eq.~(\ref{eq:cterm}) for $m=0.3$ GeV.
The solid line is the total result for $m=0.3$ GeV, corresponding to the sum of the dashed-dotted and dotted lines.}
  \label{fig:DiFF}
\end{figure}

The results for $\widetilde{D}^{\sphericalangle}_{ot}$ divided by the unpolarized DiFF $D_{1,oo}$ are plotted in Fig.~\ref{fig:DiFF}. The left panel depicts the $z$-dependent of the ratio ($M_h$ is integrated out in the region $0.3<M_h<1.6$), and the right panel depicts the $M_h$-dependence of the ratio ($z$ is integrated out over the region $0.2<z<0.9$).
The dashed line corresponds to the result for $m=0$ GeV.
The dashed-dotted line corresponds to the result from the Im$(F^{s*} F^p)$ term for $m=0.3$ GeV, while the dotted line corresponds to the nonzero result from the $C$ term in Eq.~(\ref{eq:cterm}) for $m=0.3$ GeV.
The solid line is the total result for $m=0.3$ GeV, corresponding to the sum of the dashed-dotted and dotted lines.
We find that ${\widetilde{D}^{\sphericalangle}_{ot}}/{D_{1,oo}}$ is positive in the entire $z$ and $M_h$ region when $M_h$ or $z$ is integrated out, respectively.
We note that the $\cos R_\phi$ asymmetry in the unpolarized SIDIS is proportional to the ratio $\widetilde{D}^{\sphericalangle}/D_1$  as suggested by Eq.~(44) in Ref.~\cite{Bacchetta:2003vn}, therefore, the curve in Fig.~\ref{fig:DiFF} can be also viewed as an approximate result for the $\cos \phi_R$ asymmetry in the unpolarized SIDIS.
We also observe that two difference choices on the quark mass $m$ lead to a $10 \%$ difference on the ratio ${\widetilde{D}^{\sphericalangle}_{ot}}/{D_{1,oo}}$

Using Eq.~(\ref{eq:AcosphiR}), one can express the double longitudinally polarized asymmetry $A^{\cos\phi_R}_{LL}$ in SIDIS as functions of $x$, $z$ and $M_h$ as follow
\begin{align}
\label{Ax}
A^{\cos\phi_R}_{LL}(x)=-\frac{\int dz \int dM_h 2M_h \frac{|\vec{R}|}{Q}
\frac{1}{z} (4 g_1^u(x)+g_1^d(x))\widetilde{D}^{\sphericalangle}_{ot}(z,M_h^2)
}{\int dz \int dM_h 2M_h (4 f_1^u(x)+f_1^d(x))D_{1,oo}(z,M_h^2)},
\end{align}
\begin{align}
\label{Az}
A^{\cos\phi_R}_{LL}(z)=-\frac{\int dx\int dM_h 2M_h \frac{|\vec{R}|}{Q}
 \frac{1}{z} (4 g_1^u(x)+g_1^d(x))\widetilde{D}^{\sphericalangle}_{ot}(z,M_h^2)
}{\int dx \int dM_h 2M_h (4 f_1^u(x)+f_1^d(x))D_{1,oo}(z,M_h^2)},
\end{align}

\begin{align}
\label{Amh}
A^{\cos\phi_R}_{LL}(M_h)=-\frac{\int dx\int dz \frac{|\vec{R}|}{Q}
 \frac{1}{z} (4 g_1^u(x)+g_1^d(x))\widetilde{D}^{\sphericalangle}_{ot}(z,M_h^2)
}{\int dx\int dz (4 f_1^u(x)+f_1^d(x))D_{1,oo}(z,M_h^2)}.
\end{align}

For the unpolarized distribution $f_1$ and the helicity distribution $g_1$, we also apply a spectator model result from Ref.~\cite{Bacchetta:2008af} for consistency.
As the scale dependence of the DiFF $\widetilde{D}^{\sphericalangle}$ still remains unknown, we assume that $\widetilde{D}^{\sphericalangle}$ follows the same evolution as that of the DiFF $D_1$~\cite{Ceccopieri:2007ip,deFlorian:2003cg} in leading order.
For the PDFs $f_1$ and $g_1$, we adopt the leading-order DGLAP evolution.
We admit that the scale dependence of $\widetilde{D}^{\sphericalangle}$ might be more complicated than that of $D_1$ because $\widetilde{D}^{\sphericalangle}$ is a twist-3 object.
Since here the asymmetry is the ratio between $g_1 \otimes \widetilde{D}^{\sphericalangle}$ and $f_1\otimes {D}_1$, we supposed the evolution effect will not influence the results qualitatively.

The COMPASS Collaboration~\cite{S.Sirtl} is measuring the above $A_{LL}^{\cos\phi}$ asymmetry using a 160 GeV or a 190 GeV longitudinally polarized muon beam on a longitudinally polarized nucleon target.
Using the numerical result for $\widetilde{D}^{\sphericalangle}_{ot}$, we estimate the asymmetry $A_{LL}^{\cos\phi_R}$ at the kinematical region of COMPASS
\begin{align}
&0.003<x<0.4,\quad 0.1<y<0.9, \quad 0.2<z<0.9,\nonumber\\
&0.3\mathrm{GeV}<M_h<1.6\mathrm{GeV}, \quad 1\mathrm{GeV}<Q^2, \quad W>5\mathrm{GeV}.
\label{eq:cuts}
\end{align}
Here $W$ is invariant mass of the virtual photon-nucleon system. The kinematical cuts in
Eq.~(\ref{eq:cuts}) are implemented in the calculation numerically by
imposing those constraints in the computation code.

\begin{figure}
  \centering
  \includegraphics[width=0.32\columnwidth]{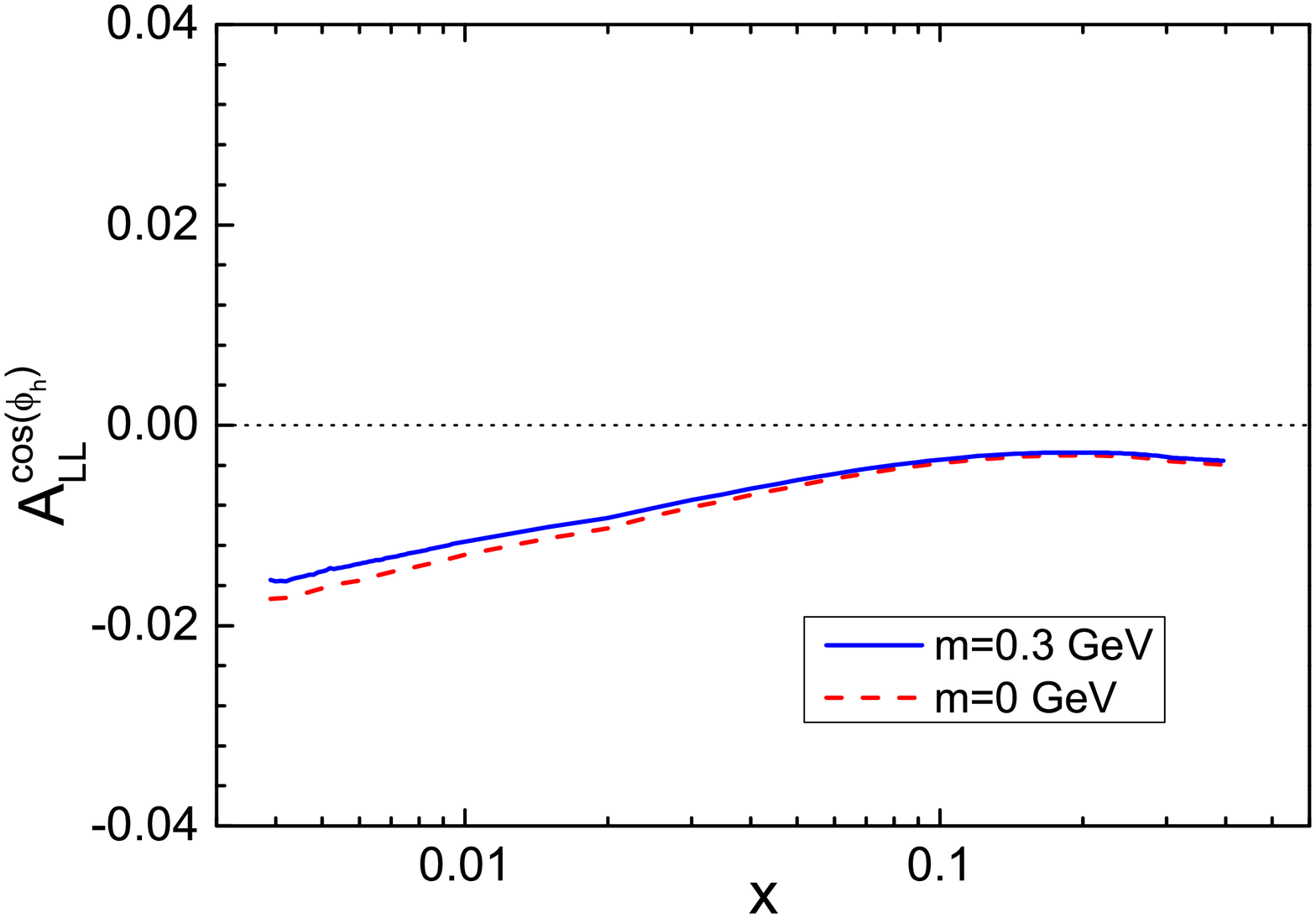}
  \includegraphics[width=0.32\columnwidth]{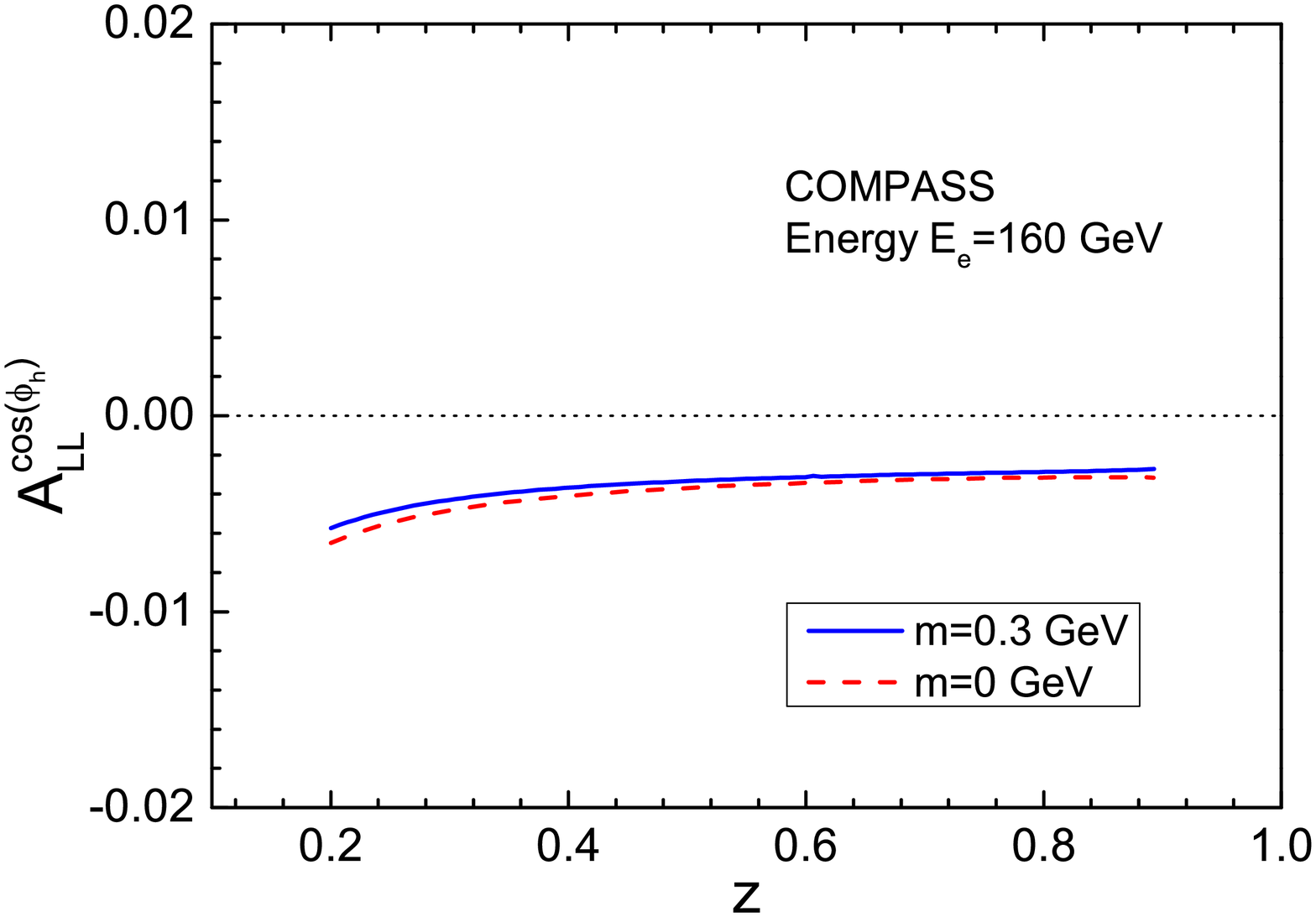}
  \includegraphics[width=0.32\columnwidth]{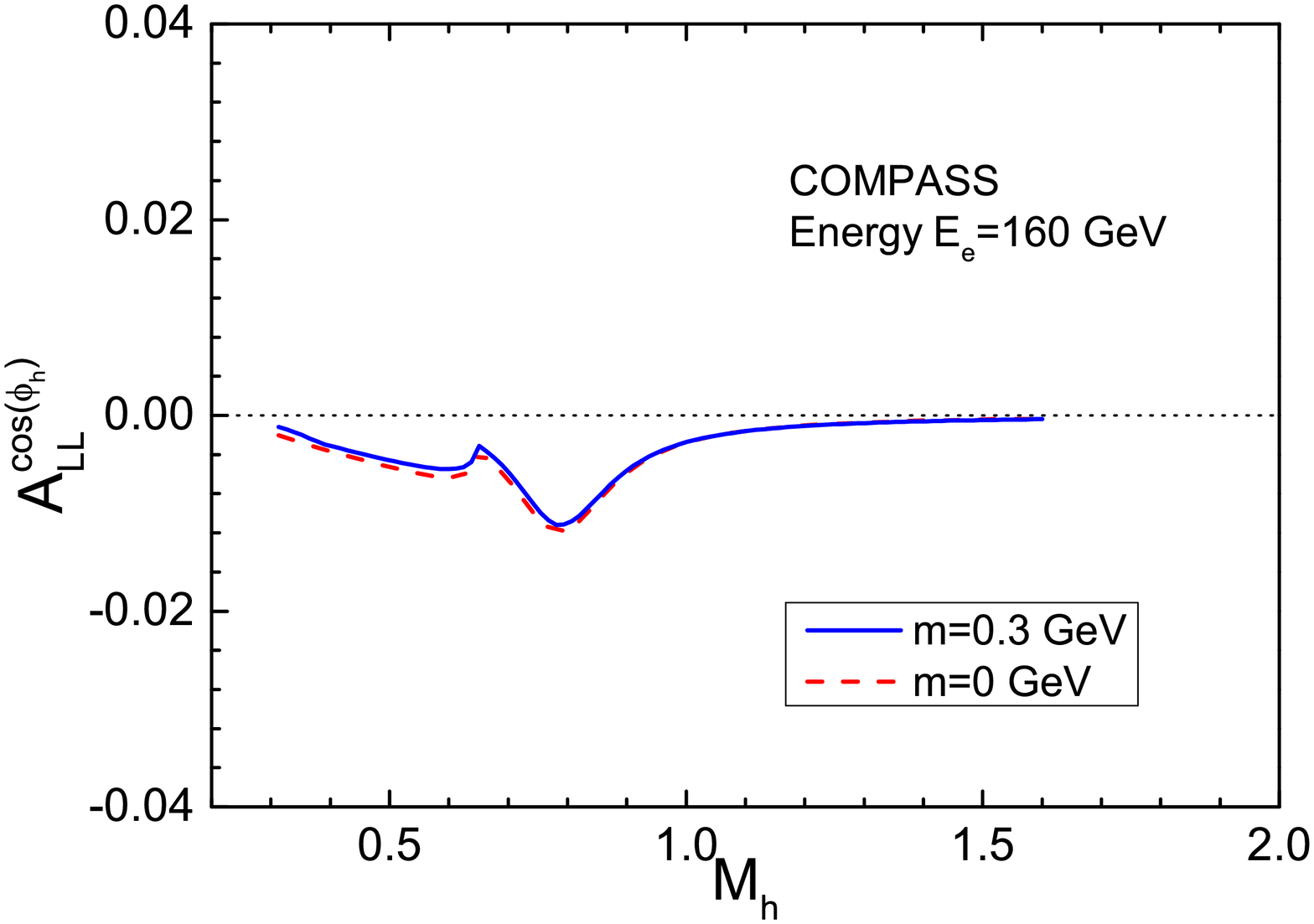}
  \caption{The $\cos\phi_R$ azimuthal asymmetry in dihadron production off the longitudinally polarized proton as functions of $x$ (left panel), $z$ (central panel) and $M_h$ (right panel) at COMPASS.
  The dashed and solid lines corresponds to the results from $m=0$ GeV and $m=0.3$ GeV.}
  \label{fig:asy}
\end{figure}

In the left, central, and right panels of Fig.~\ref{fig:asy}, we plot the $x$, $z$ and $M_h$ dependent $\cos\phi_R$ asymmetry at the COMPASS kinematics.
The solid and dashed lines correspond to the results for $m=0.3$ GeV ad $m=0$ GeV, respectively.
Again, the difference between this two results is about $10\%$.
We find that the asymmetry is negative due to the minus sign in Eq.~(\ref{eq:AcosphiR}).
The asymmetry roughly decreases with increasing $x$.
The estimated magnitude of the asymmetry at COMPASS is about 0.01, which is smaller than the $\sin\phi_R$ asymmetry in single longitudinally polarized SIDIS~\cite{S.Sirtl,Yang:2019aan}.

\begin{figure*}
  \centering
  \includegraphics[width=0.32\columnwidth]{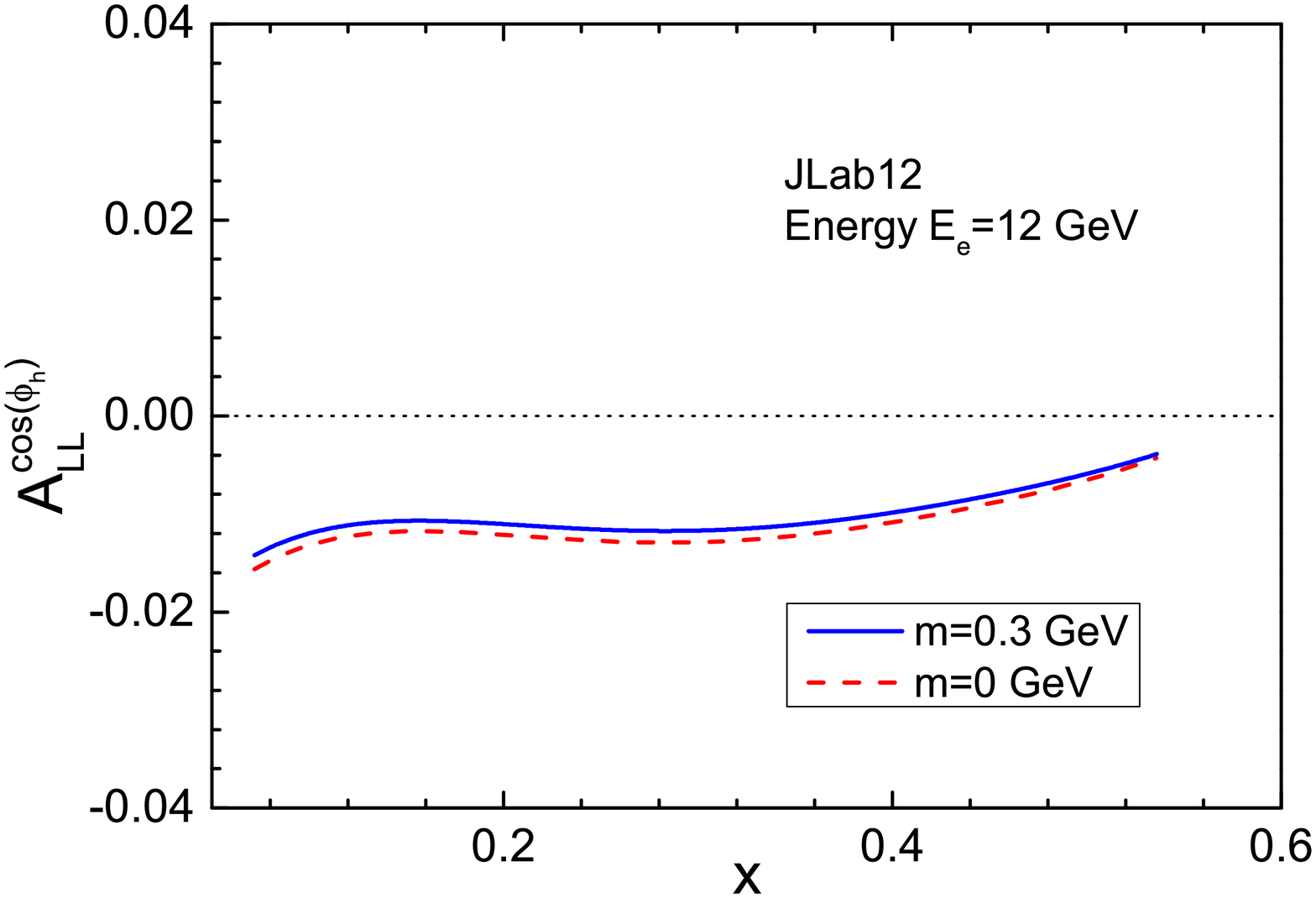}
  \includegraphics[width=0.32\columnwidth]{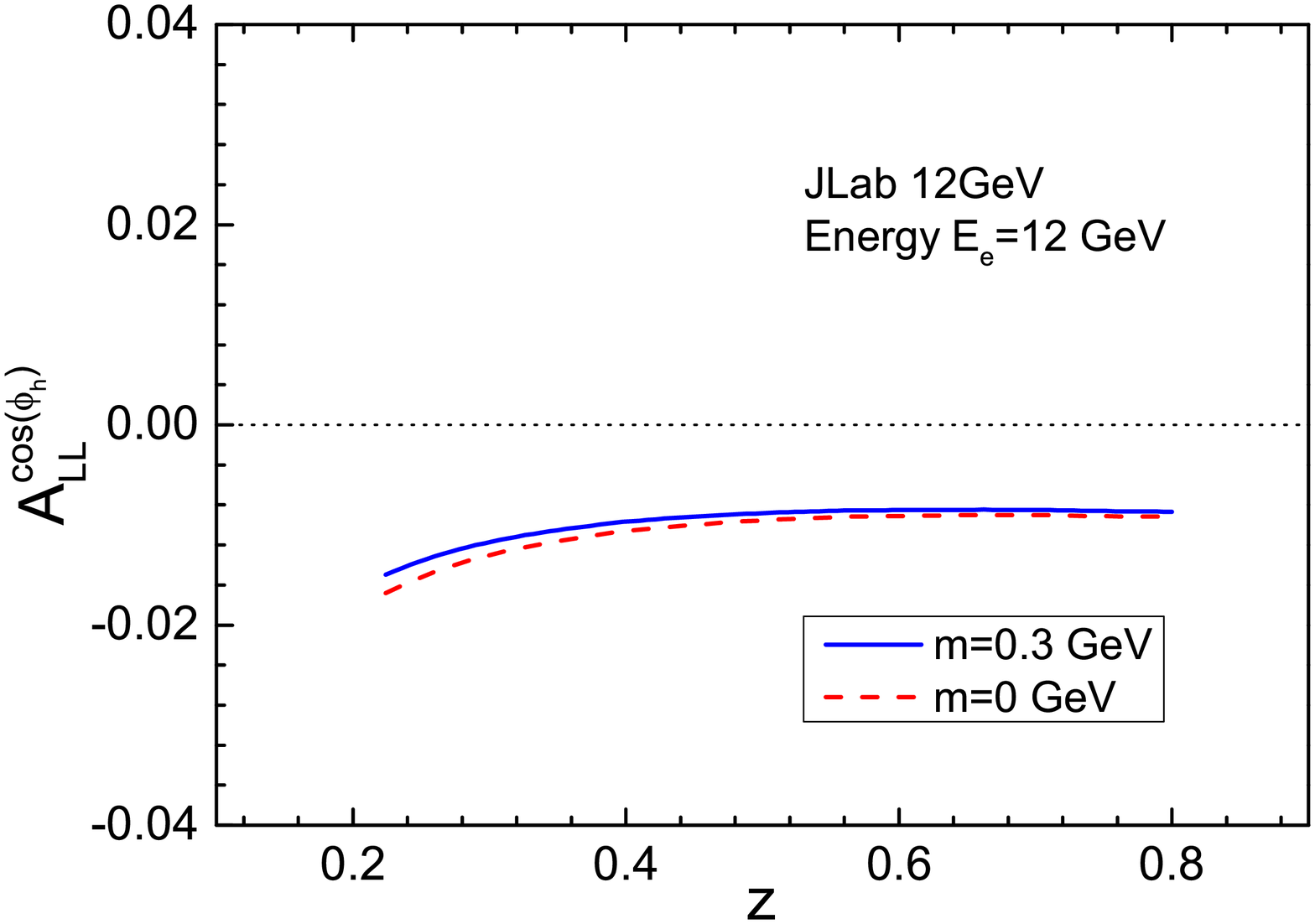}
  \includegraphics[width=0.32\columnwidth]{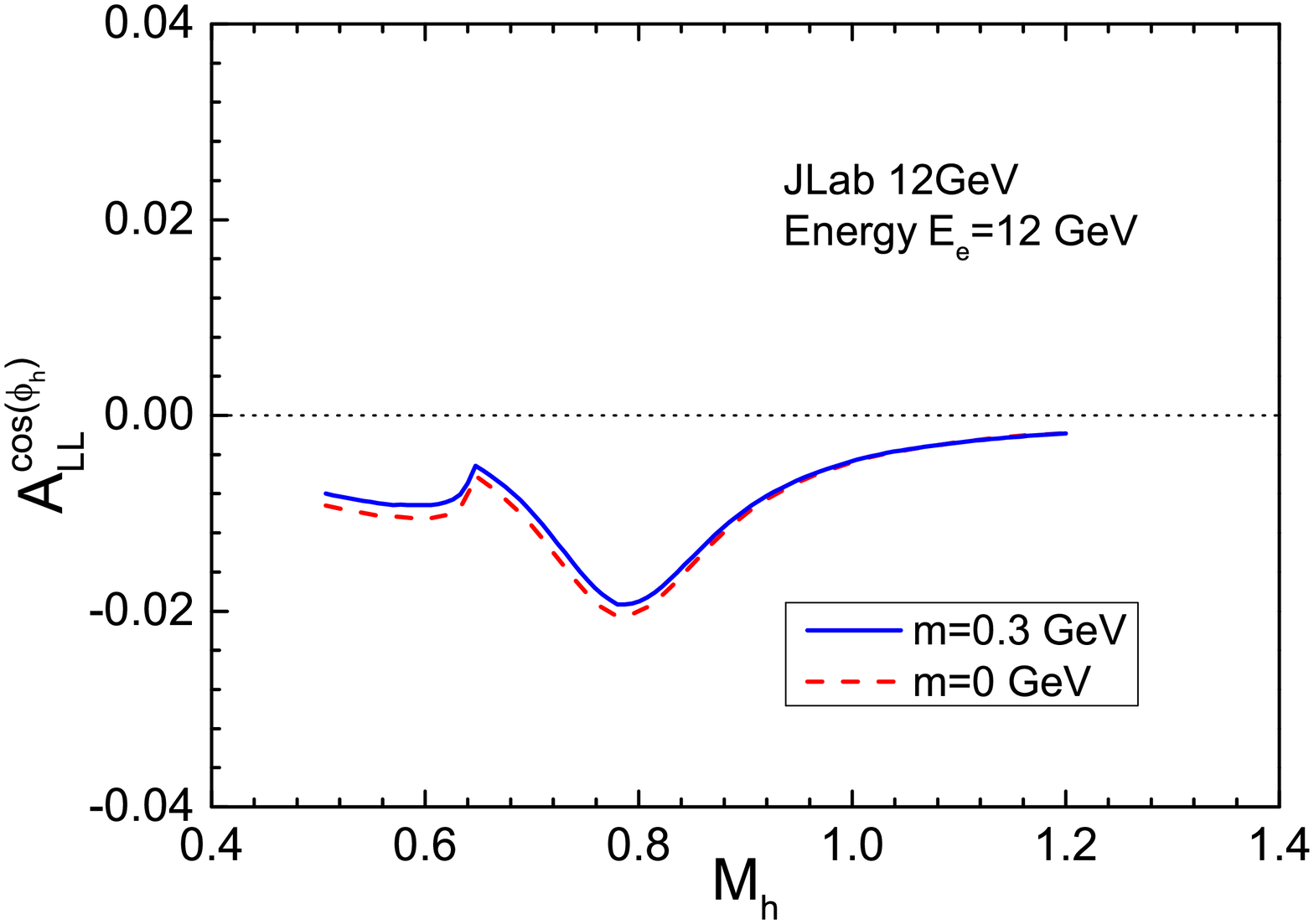}
  \caption{The $\cos\phi_R$ azimuthal asymmetry in dihadron production off the longitudinally polarized proton as functions of $x$ (left panel), $z$ (central panel) and $M_h$ (right panel) at JLab 12GeV. The dashed and solid lines corresponds to the results from $m=0$ GeV and $m=0.3$ GeV.}
  \label{fig:jlab12}
\end{figure*}

For comparison, we also predict the asymmetry $A_{LL}^{\cos\phi_R}$ at JLab 12GeV using the following kinematical configuration
\begin{align}
&\quad 0.072<x<0.532,\quad 0.2<y<0.95,\quad  0.2<z<0.8,\quad 0.5 \textrm{GeV} <M_h<1.2\textrm{GeV},\nonumber\\
&\quad  E_e = 12\mathrm{GeV},\quad  W>4 \mathrm{GeV}^2,\quad  1<Q^2<6.3 \mathrm{GeV}^2.
\label{eq:cuts2}
\end{align}
The corresponding results are shown in Fig.~\ref{fig:jlab12}.
In general, The asymmetry at JLab 12GeV is found to be larger than that at COMPASS because the $A_{LL}^{\cos\phi_R}$ asymmetry is a twist-3 effect and the c.m. energy at JLab is smaller.

\begin{figure*}
  \centering
  \includegraphics[width=0.32\columnwidth]{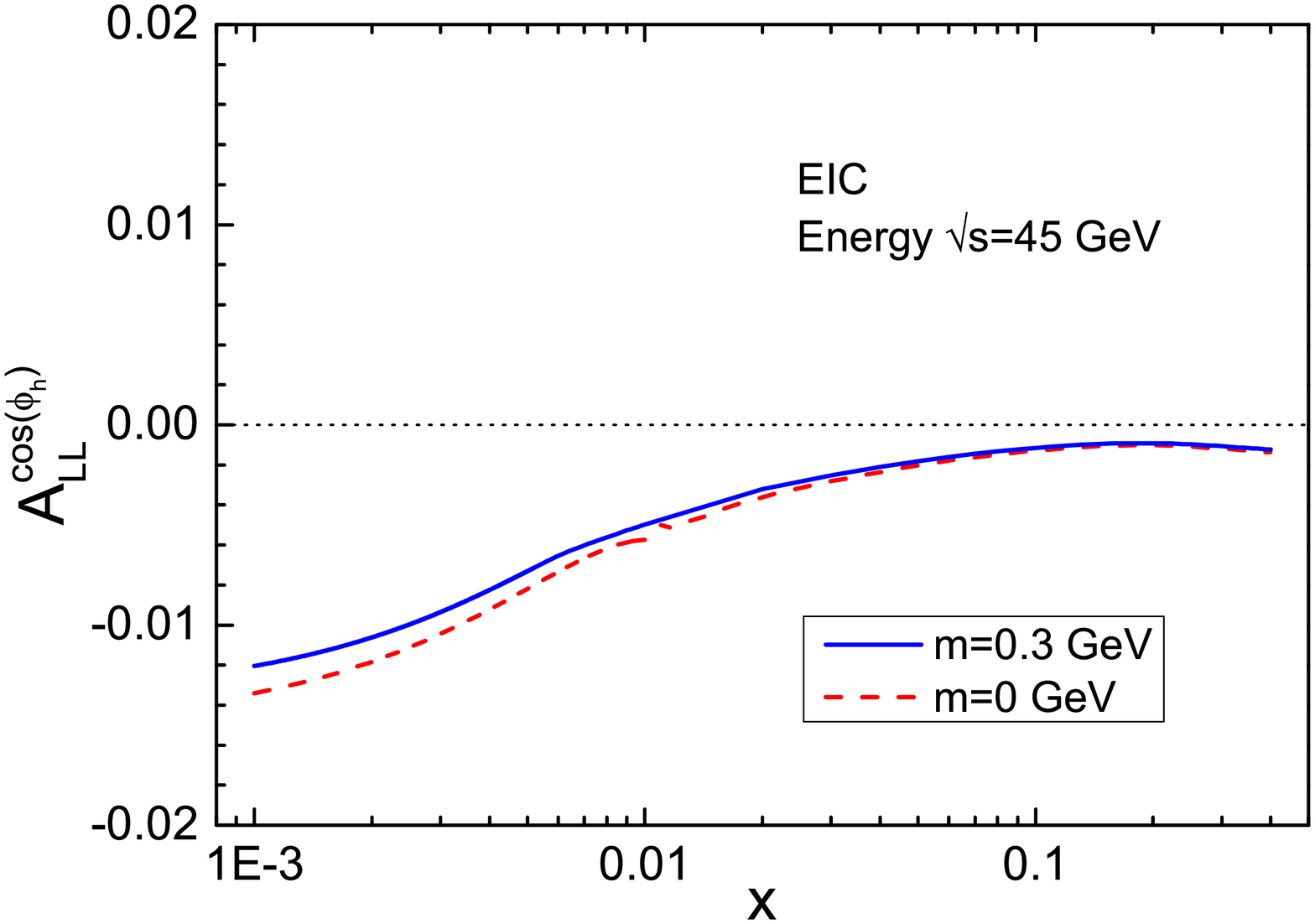}
  \includegraphics[width=0.32\columnwidth]{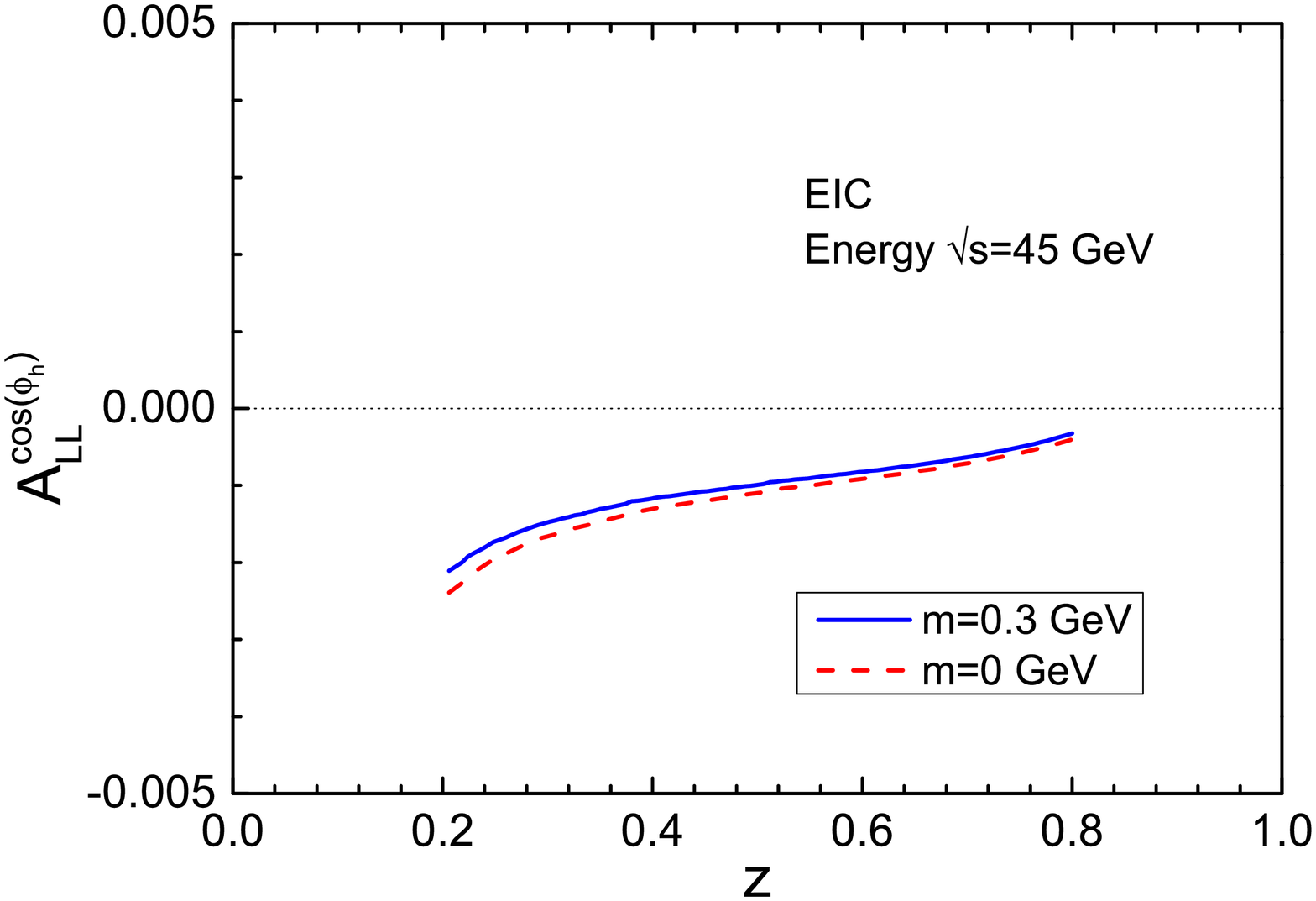}
  \includegraphics[width=0.32\columnwidth]{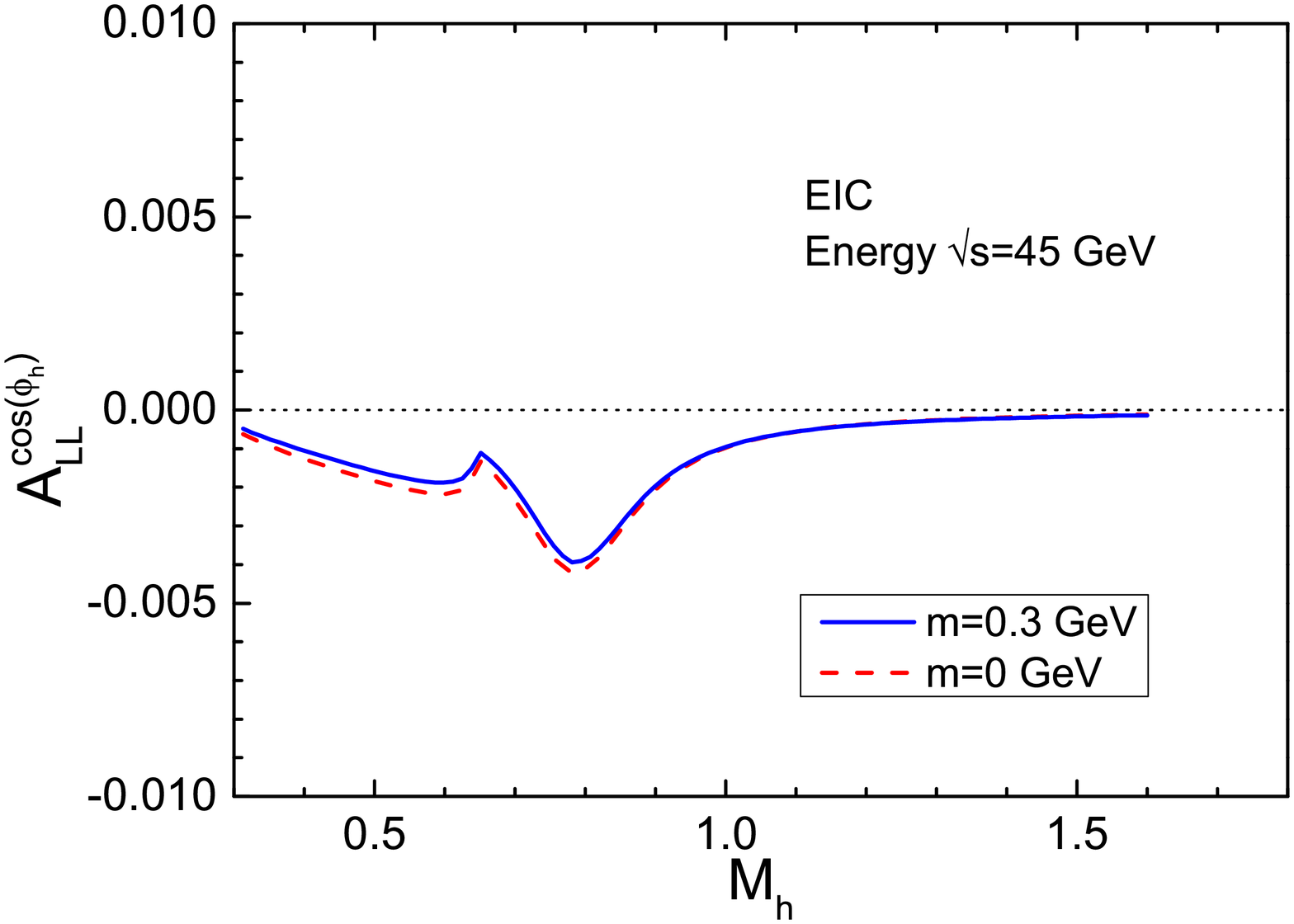}
  \caption{The $\cos\phi_R$ azimuthal asymmetry in dihadron production off the longitudinally polarized proton as functions of $x$ (left panel), $z$ (central panel) and $M_h$ (right panel) at an EIC. The dashed and solid lines corresponds to the results from $m=0$ GeV and $m=0.3$ GeV.}
  \label{fig:eic}
\end{figure*}

We also present the $\cos\phi_R$ asymmetry in double polarized SIDIS at the kinematical configuration of a future EIC facility~\cite{Matevosyan:2015gwa}:
\begin{align}
&\sqrt{s}=45\mathrm{GeV}, \quad 0.001<x<0.4,\quad 0.01<y<0.95, \quad 0.2<z<0.8,\nonumber\\
&0.3 \textrm{GeV} <M_h<1.6\textrm{GeV},\quad Q^2>1\mathrm{GeV}^2,\quad  W>5\mathrm{GeV}.
\end{align}
The asymmetry vs $x$, $z$ and $M_h$ are plotted in the left, central and right panels of Fig.~\ref{fig:eic}.
We find that the overall tendency of the asymmetry at the EIC is similar to that at COMPASS.
Although the size of the asymmetry is smaller due to the higher-twist nature of the asymmetry, it is still measurable at the kinematics of EIC.

Finally, we note that a recent measurement on the single-transverse spin asymmetry of dihadron production in SIDIS by COMPASS~\cite{Adolph:2014fjw} shows that the asymmetry is similar to the Collins asymmetry of single hadron production in SIDIS.
This hints that the spin-dependent DiFFs could be generated by the single-hadron spin-dependent fragmentation functions.
The closed relation between the Collins fragmentation function $H_1^\perp$ and the spin-dependent DiFF $H_1^{\sphericalangle}$ in large $M_h$ region has also been suggested in Ref.~\cite{Zhou:2011ba}, which showed that
both the leading twist unpolarized dihadron fragmentation function and the leading twist interference FF (IFF) can be perturbatively computed in terms of the normal single hadron fragmentation function when the invariant mass is sufficiently large.
Therefore, there is the possibility that the similar analysis can be applied to the twist-3 IFF case as well.
On the other hand, a calculation of SSA in dihadron production based on the spectator model result~\cite{Bacchetta:2006un} for $H_1^{\sphericalangle}$ can also well describe the COMPASS data (see solid blue lines in Fig.~5 of Ref.~\cite{Adolph:2014fjw}) for the $x$ and $z$ dependent shape, as well as for the $M_h$ dependent shape in the $\rho$ meson region.
This may suggest that the resonance interference mechanism applied in our paper can also responsible for the spin asymmetry in dihadron production.
Due to the lack of further data, at the moment it is hard to say which mechanism should be preferred.
Further theoretical and experimental studies are needed in order to discriminate different mechanisms for the spin-dependent DiFFs.

\section{Conclusion}
\label{Sec.conclusion}

In this work, we studied the origin of the $\cos\phi_R$ asymmetry of hadron pair production in double polarized SIDIS: $l^\rightarrow+p^\rightarrow \rightarrow h_1+h_2 +X$
The asymmetry can originate from the coupling of the twist-3 DiFF $\widetilde{D}^{\sphericalangle}$ and the helicity distribution $g_1(x)$. Another potential contribution, the coupling $e_L(x)\, H_1^{\sphericalangle}$, will not give rise to the asymmetry because of the time-reversal invariant constraint $\int d^2p_T\,e_{L}(x,p^2_{T})=0$.
We applied a spectator model for the quark-gluon-quark correlator and calculated the $s$-wave and $p$-wave interference DiFF $\widetilde{D}^{\sphericalangle}_{ot}$, the leading term of $\widetilde{D}^{\sphericalangle}$ in the partial wave expansion.
Using the numerical results for $\widetilde{D}^{\sphericalangle}_{ot}$, we estimated the $\cos \phi_R$ asymmetry as functions of $x$, $z$ and $M_h$ at the kinematics of COMPASS, JLab 12GeV, and EIC.
We found that the asymmetry at COMPASS and JLab12 is about 1-2 percent and may be measurable.
Therefore, the measurement of the $\cos\phi_R$ asymmetry at COMPASS and JLab12 may provide unambiguous information on $\widetilde{D}^{\sphericalangle}$.

\section{Acknowledgements}

This work is partially supported by the NSFC (China) grant 11575043.  W. Y and H. L contributed equally to this work and should be considered as co-first authors.

\end{document}